\documentclass[12pt,preprint]{aastex}
\def\p/{\mbox{$^1$}}
\def\pp/{\mbox{$^2$}}
\def\ppp/{\mbox{$^3$}}
\def\pppp/{\mbox{$^4$}}
\def\m/{\mbox{$^{-1}$}}
\def\mm/{\mbox{$^{-2}$}}
\def\mmm/{\mbox{$^{-3}$}}
\def\mmmm/{\mbox{$^{-4}$}}
\def\Ms/{\mbox{M$_\odot$}}

\shorttitle{Numerical biases on IMF determinations created by binning}
\shortauthors{Ma\'{\i}z Apell\'aniz \& \'Ubeda}

\slugcomment{Submitted for publication in the Astrophysical Journal}

\begin{document}

\title{Numerical biases on IMF determinations created by binning}

\author{J. Ma\'{\i}z Apell\'aniz\altaffilmark{1,2} and L. \'Ubeda}
\affil{Space Telescope Science Institute\altaffilmark{3}, 3700 San Martin 
Drive, Baltimore, MD 21218, U.S.A.}

% ---------------affiliations of the science group -----------------------

\altaffiltext{1}{Affiliated with the Space Telescope Division of the European 
Space Agency, ESTEC, Noordwijk, Netherlands.}
\altaffiltext{2}{e-mail contact: {\tt jmaiz@stsci.edu}.}
\altaffiltext{3}{The Space Telescope Science Institute is operated by the
Association of Universities for Research in Astronomy, Inc. under NASA
contract No. NAS5-26555.}

\begin{abstract}
We detect and quantify significant numerical biases in the determination of the slope of power laws 
with Salpeter (or similar) indices from uniformly-binned data using chi-square minimization. The biases 
are caused by the correlation between the number of stars per bin and the assigned weights and are especially 
important when the number of stars per bin is small. This result implies the existence of systematic
errors in the values of IMFs calculated in this way. We propose as an alternative using variable-size bins
and dividing the stars evenly among them. Such variable-size bins yield very small biases 
that are only weakly dependent on the number of stars per bin. Furthermore, we show that they allow for the 
calculation of reliable IMFs with only a small total number of stars. Therefore, they are a preferred 
alternative to the standard uniform-size binning.
\end{abstract}


\section{Introduction}

	The data explosion of the last decade created by the growth of telescope diameters, number of space
missions, and detector surface area has produced a comparable growth in the amount of objects that can be
observed with a single exposure or analyzed in a single paper. In the field of stellar astronomy, this has
translated into thousands or larger amounts of stars being included in a single color-magnitude diagram or a
single luminosity or mass function. Indeed, many articles aim at measuring the initial mass function (or IMF)
of a cluster or of a stellar
population by measuring the luminosity of each star, converting the results into masses (using isochrone 
fitting or directly counting the number of stars between evolutionary tracks), and fitting a power 
law or similar function to the masses organized in bins using chi-square minimization. 

	A known numerical effect takes place when data are binned and a function is fitted to the 
outcome: a bias in the derived parameters can be (and usually is) present if there is a low number of objects 
in some of the bins (see, e.g. \citealt{BeviRobi92,NousShue89}). The problem originates in the strong 
anticorrelation between the data and the weights in chi-square minimization for data with
Poisson uncertainties (\citealt{Wheaetal95}, the same should be true for binomial uncertainties, which in
most cases can be well approximated by Poisson ones, see e.g. \citealt{BeviRobi92}). 
The bias can be especially large when
some bins have 0 or 1 counts; in the former case some algorithms set the weights to correspond to that of 1
count, but that does not eliminate the effect. Furthermore, the bias can also be present 
when the number of objects per bin is not too small if some bins are more heavily populated than others
\citep{Wheaetal95}. The problems associated with this bias 
are usually correctly dealt with in some astronomical fields, such as in high-energy astronomy,
but it is surprising to find that many of the articles on the IMF appear to ignore them. Thus, it is not strange 
to find articles in which an IMF is calculated with some bins having just a few stars and other bins with tens 
or hundreds of them (see \citealt{HumpMcEl84,Massetal89,Bejaetal01} for some examples). Such a binning scheme
introduces biases in the calculated IMF slopes \citep{Krou01,Elme04}.

	There are different approaches to minimizing or eliminating binning biases. \citet{Kearetal95} suggest
using the fitted number of counts (instead of the real number of counts) to calculate the weight of each bin, a
method that is iterative by nature since one does not know the fitted number of counts a priori. An
alternative recommended by \citet{DAgoStep86} involves two measures:

\begin{itemize}
  \item Define bins of variable size in $x$ and adjust them in such a way that each one of them has
	approximately the same number of objects. The reasoning behind this idea is to assign the same
	statistical weight to each bin and thus minimize biases.  
  \item If $N$ is the total number of objects, then divide the data into $\approx 2\cdot N^{2/5}$ bins.
\end{itemize}

	In this article we set out to quantify the importance of binning biases for the determination of the
IMF by means of a series of simple numerical experiments. We point out that our results should be relevant 
not only to IMF calculations but to similar problems as well, such as the calculation of cluster 
mass or luminosity distributions. We will start by using ``standard'' uniform-size bins 
and then we will explore the use of non-uniform ones, as suggested by \citet{DAgoStep86}.

\section{Definitions and experiments}

	Following the \citet{Scal86} notation, we define the IMF for stars $f(m)$ such 
that $f(m)\cdot dm$ is the number of stars formed at the same time in some volume of space with mass in the 
interval $m$ to $m+dm$. Throughout this article, we will assume that the true IMF follows a power-law
distribution:

\begin{equation}
\frac{dn}{dm} = f(m) = A \cdot m^{\gamma},
\label{imf1}
\end{equation}

\noindent where the Salpeter slope is given by $\gamma = -2.35$ \citep{Salp55}.

	The number of stars created in the interval $[m_a, m_b]$ can be obtained by integrating 
Eqn.~\ref{imf1}: 

\begin{equation}
\int_{m_a}^{m_b} \frac{dn}{dm} dm = A \cdot \int_{m_a}^{m_b} m^{\gamma} dm
\label{imf2}
\end{equation}

\noindent to obtain:

\begin{equation}
n\left|_{m_a}^{m_b}\right. = 
 n(m_b) - n(m_a) = \frac{A}{\gamma + 1} \cdot \left[ {m_b}^{\gamma + 1}- {m_a}^{\gamma + 1} \right]  
 \hspace{5mm}; \gamma \neq -1.
\label{imf3}
\end{equation}

	Defining $\Delta m_i = m_b - m_a$  and $x_i= \frac{m_a+m_b}{2}$ it is possible to rewrite the last 
expression as:

\begin{equation}
N_i \equiv n\left|_{m_a}^{m_b}\right.  = 
 \frac{A}{\gamma + 1} \cdot \left[   \left( x_i + \frac{\Delta m_i}{2} \right)^{\gamma + 1}-  
 \left( x_i-\frac{\Delta m_i}{2} \right)^{\gamma + 1} \right].
\label{imf4}
\end{equation}

	Finally, we obtain the logarithm of both sides to obtain an expression in the form of 
$y_i$ (logarithm of the number of stars in bin $i$) as a function of $x_i$ (mass at the center of bin $i$ as 
defined in a linear scale):

\begin{equation}
y_i \equiv \log_{10} N_i =
\log_{10} \left( \frac{A}{\gamma + 1}  \cdot \left[   \left( x_i + \frac{\Delta m_i}{2} \right)^{\gamma + 1}-  
\left( x_i-\frac{\Delta m_i}{2} \right)^{\gamma + 1} \right]  \right).
\label{imf5}
\end{equation}

	For our numerical experiments we used a random number generator to produce 1000 realizations with
1000 stars each, distributed according to Eqn.~\ref{imf1} with a Salpeter slope between $m$ = 6.31 M$_\odot$ 
($\log_{10} m/$M$_\odot = 0.8$) and $m$ = 158.49 M$_\odot$ ($\log_{10} m/$M$_\odot = 2.2$). The extremes were
selected as typical for studies of massive stars (see, e.g. \citealt{OeyClar05}), 
where the IMF has been measured to be close to the Salpeter
value under different circumstances, but the conclusions of this work are not expected to change if somewhat 
different values are used. The 1000 realizations were manipulated in two different ways to test different 
conditions:

\begin{itemize}
  \item Selecting the first 30, 100, 300, and 1000 (all) stars to simulate different total number of stars
	$N=\sum N_i$.
  \item Binning the data into 3, 5, 10, 30, or 50 bins to test whether there is an optimal bin size. Given
	that the \citet{DAgoStep86} recommendation yields 7.8, 12.6, 19.6, and 31.7 bins for our four
	values of $N$, the bins selected here should be able to test its validity for our experiments.
\end{itemize}

	We performed experiments using three types of binning:

\begin{enumerate}
  \item Uniform bin size in logarithmic scale (or $\log_{10}[(x_i+\Delta m_i/2)/(x_i-\Delta m_i/2)]$ constant) with 
	the left edge of the first bin, $m_{\rm down}$, equal to $10^{0.8}$~M$_\odot$ and the right edge of the last 
	bin, $m_{\rm up}$, equal to $10^{2.2}$~M$_\odot$ (i.e. the input values used for the random generation of all 
	realizations).
  \item Approximately constant number of stars in each of the bins with $m_{\rm down} = 10^{0.8}$~M$_\odot$ and 
	$m_{\rm up} = 10^{2.2}$~M$_\odot$. This rule cannot be made exact since it is not possible to divide e.g. 
	100 stars in 30 bins with the same number of stars in each. In such a case, we design our bins so that they
	contain e.g. 3, 4, 3, 3, 4, 3\ldots\ stars.
  \item Same as the previous experiment but with $m_{\rm down}$ and $m_{\rm up}$ determined from the data in each
	realization.
\end{enumerate}

	The issue of the weights to be applied to the problem of IMF fitting has generated some confusion. Some 
authors have even decided to skip it altogether by not using any \citep{Massetal95b}, but such a strategy can
yield large biases. When calculating an IMF we are assuming that an underlying physical law determines the true 
probability distribution $f(m)$, which we could measure with arbitrary precision if an infinite number of stars 
were generated from it. In reality, we are limited to analyze 
finite samples of size $N$ drawn from $f(m)$. It is easy to show that, under such circumstances, the 
value of $N_i$ follows a binomial distribution characterized by $N$ and $p_i$, where:

\begin{equation}
p_i = \frac{\int_{x_i-\Delta m_i/2}^{x_i+\Delta m_i/2} f(m)\, dm}{\int_{m_{\rm down}}^{m_{\rm up}} f(m)\, dm},
\end{equation}

\noindent which has a mean of $Np_i$ and a variance of $Np_i(1-p_i)$. Note that the Poisson approximation corresponds to 
the case where $p_i \ll 1$; however, if a bin contains a large fraction of the objects, then such approximation is no
longer correct. The problem here, as noted by \citet{Wheaetal95}, is that the true $p_i$ needs to be determined 
from the true $f(m)$, which is unknown! One obvious alternative is to substitute $p_i$ by $N_i/N$, the value derived
from the data, so that the estimate for the uncertainty associated to $N_i$ becomes $\sqrt{N_i(N-N_i)/N}$. That is the
alternative we will use, keeping in mind that this step is precisely the source of the potential biases we have
previuosly referred to \citep{Wheaetal95} and which we are trying to minimize by selecting bins with similar numbers of 
stars.

	With the above considerations and given that (a) $y_i=\log_{10} N_i$, and (b) the weight $w_i$ associated to 
a value $y_i$ with uncertainty $s_i$ (i.e. $y_i\pm s_i$)\footnote{$s_i$ is the standard deviation derived from
the parent distribution of $y_i$ and its measured value, not from the expected one. Therefore, what we are describing 
here is the modified chi-square minimization method \citep{Wheaetal95}.} when using a chi-square minimization 
algorithm is $1/s_i^2$, we have:

\begin{equation}
w_i = \frac{N_iN}{(N-N_i)(\log_{10}e)^2},
\end{equation}

\noindent which has the consequence of giving zero weight to bins with no stars. This has the advantage of 
yielding no numerical problems there, as opposed to using $N_i$ for the independent variable, which produces
infinite weights\footnote{Note, however, that in our case $w_i$ becomes infinite when $N_i=N$ but, of course, trying to 
fit a function to a histogram that has all data in a single bin is an ill-defined problem.}. 

	For each of the $k$ ($k$ = 1,1000) realizations in each of the $N$ + bin-size combinations, 
the resulting data ($x_i$, $\Delta m_i$, $y_i$, $w_i$) was fitted using a chi-square minimization algorithm 
programmed in IDL in order to obtain $\gamma_k$ and its uncertainty $\sigma_k$ (the algorithm also yields 
$A_k$ and its uncertainty, which will be ignored here).
In order to test for a possible algorithm dependence, we calculated our results using both
(a) {\tt CURVEFIT}, which uses gradient expansion, can be found in the standard IDL distribution, and is
originally based on Numerical Recipes \citep{Presetal86}; and (b) Craig B. Markwardt's\footnote{See
{\tt http://cow.physics.wisc.edu/\~{}craigm/idl/idl.html}.} {\tt MPCURVEFIT}, which is based on MINPACK-1, available
from Netlib. No significant differences were found between the two algorithms.

	We analyzed the results for the ensemble of 1000 realizations in each $N$ + bin size
combination. More specifically, we calculated the mean value for the power-law exponent:

\begin{equation}
\overline{\gamma} = \frac{1}{1000}\sum_{k=1}^{1000} \gamma_k,
\end{equation}

\noindent the mean uncertainty in the power-law exponent:

\begin{equation}
\overline{\sigma} = \frac{1}{1000}\sum_{k=1}^{1000} \sigma_k,
\end{equation}

\noindent and the bias normalized with respect to the uncertainty:

\begin{equation}
b = \frac{1}{1000}\sum_{k=1}^{1000} \frac{\gamma_k+2.35}{\sigma_k}.
\end{equation}

	The value of $b$ is the logical criterion to judge the existence of biases. If $|b|\ll 1$, then the 
fitting method will be unbiased because it will yield values that will be larger than the real one on
$\approx$50\% of the occasions and smaller on another $\approx50$\%. If, on the other hand, $|b|\sim 1$ or 
larger, a significant bias will exist. 

	Of course, we want to test whether our experiments yield biased or unbiased results in order to decide which
binning technique is the optimum one. However, a fitting method can be unbiased but still yield an
incorrect uncertainty estimate. A more complete test would be to analyze the distribution of the quantity 
$(\gamma_k+2.35)/\sigma_k$, whose mean is given by $b$ and whose standard deviation is given by:

\begin{equation}
\beta = \sqrt{\frac{1}{1000}\sum_{i=1}^{1000} \left(\frac{\gamma_k+2.35}{\sigma_k}-b\right)^2}.
\end{equation}

	A binning technique that is both unbiased and yields correct uncertainty estimates should produce a 
distribution for $(\gamma_k+2.35)/\sigma_k$ that resembles a Gaussian with $b=0$ and $\beta=1$. Under such 
circumstances, we can predict that the true value of the slope of the IMF will be e.g. within $\gamma_k-\sigma_k$ 
and $\gamma_k$ approximately 34.1\% (or e.g. within $\gamma_k$ and $\gamma_k+2\sigma_k$ approximately 47.7\%) 
of the times the experiment is executed. If $\beta$ is significantly different from 1 but $b\approx 0$, then the
technique will yield unbiased results with incorrect uncertainty estimates. If, on the other hand, 
$\beta\approx 1$ but $b$ is significantly different from zero, the technique will be biased (e.g. will have a
systematic error) but its uncertainty estimate (sometimes called the random error) will be correct, as previously
mentioned. Of course, the worst case scenario implies values of $b$ and $\beta$ significantly different from zero and one,
respectively, in which case the technique produces both systematic errors and incorrect uncertainty estimates.

	Finally, we expect the results from different experiments to depend on the existence of 
bins with few or no stars. In order to analyze that effect we define $\overline{N}_{i,{\rm min}}$ to be the 
mean $N_i$ in the bin with the lowest number of counts, which for uniform-size bins and a Salpeter power law 
will be the rightmost one. For our variable-size bin experiments, $\overline{N}_{i}$ is the same for any bin so 
there is no need to select an specific one.

\section{Experiment 1: Uniform-size bins}

\begin{deluxetable}{rrrrrrrrrrrrrrrrrrr}
\rotate
\tablecaption{Results for the uniform bin-size case for 3, 5, 10, 30, and 50 bins.}
\tablewidth{0pt}
\tabletypesize{\footnotesize}
\tablehead{stars & & \multicolumn{5}{c}{$\overline{\gamma}$} & 
                   & \multicolumn{5}{c}{$\overline{\sigma}$} & 
                   & \multicolumn{5}{c}{$b$                } \\
                 & & \colhead{3} & \colhead{5} & \colhead{10} & \colhead{30} & \colhead{50} &
                   & \colhead{3} & \colhead{5} & \colhead{10} & \colhead{30} & \colhead{50} &
                   & \colhead{3} & \colhead{5} & \colhead{10} & \colhead{30} & \colhead{50}}
\startdata
   30   & &
-2.308  & -2.207  & -2.050  & -1.691  & -1.513  & &    
 0.303  &  0.275  &  0.282  &  0.293  &  0.296  & &    
 0.376  &  0.655  &  1.181  &  2.393  &  2.986  \\     
   100  & &
-2.343  & -2.306  & -2.244  & -2.042  & -1.895  & &    
 0.155  &  0.144  &  0.148  &  0.154  &  0.156  & &    
 0.176  &  0.376  &  0.772  &  2.058  &  2.988  \\     
   300  & &
-2.345  & -2.335  & -2.316  & -2.231  & -2.157  & &    
 0.088  &  0.083  &  0.085  &  0.088  &  0.089  & &    
 0.121  &  0.224  &  0.430  &  1.384  &  2.200  \\     
   1000 & &
-2.345  & -2.344  & -2.339  & -2.314  & -2.289  & &    
 0.048  &  0.045  &  0.046  &  0.048  &  0.049  & &    
 0.151  &  0.163  &  0.260  &  0.766  &  1.275  \\     
\enddata
\label{uniformbin}
\end{deluxetable}

\begin{figure}
%\centerline{\includegraphics*[width=\linewidth]{biascorr1.ps}}
\centerline{\includegraphics*[width=\linewidth]{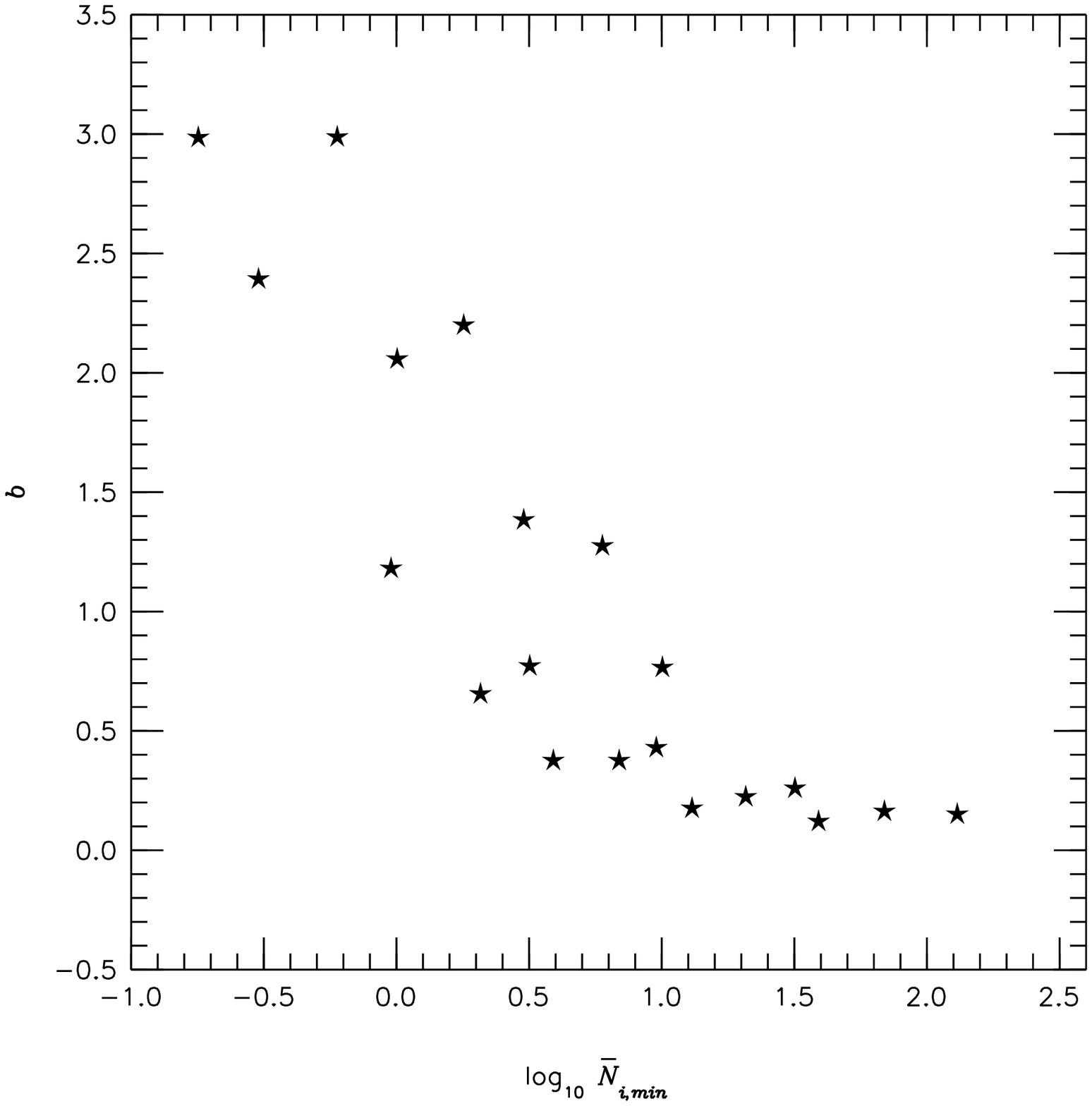}}
\caption{Bias as a function of $\overline{N}_{i,{\rm min}}$ for the first experiment. Note that
$\overline{N}_{i,{\rm min}}$ can be smaller than 1 because it is a property derived from the
parent distribution.}
\label{biascorr1}
\end{figure}

	We show in Table~\ref{uniformbin} the values of $\overline{\gamma}$, $\overline{\sigma}$,
and $b$ for the first experiment, in which we use a uniform bin size. In all of the twenty cases,
the value of $\overline{\gamma}$ is found out to be larger that $-$2.350, with the maximum at 
$-$1.513. Similarly, $b$ is always positive. In 
a few cases, the bias is small but in most of them it is quite large. $b$ is strongly anticorrelated
with $\overline{N}_{i,{\rm min}}$, as we show in Fig.~\ref{biascorr1}. This is an expected behavior,
since for bins with a small number of stars the difference between the used weight and the real 
weight (the one derived from the parent, not the sample distribution) becomes larger in relative 
terms. Furthermore, there appears to be a critical value around 
$\overline{N}_{i,{\rm min}}\approx 10$: for lower values biases are quite large while for higher 
ones they are small (though not always negligible). These results confirm what 
we mentioned in the introduction: {\it For a binning scheme where the number of objects per bin is
highly variable, significant biases can be present even if all bins have more than one star.}

	The existence of significant biases makes the use of uniform-size bins inadvisable for 
calculating IMFs. However, if no alternative is available, one should prefer those results based on
a small number of bins, since in that case biases are smaller. Increasing the number of bins reduces
the uncertainty estimate slightly but at the cost of introducing significantly larger biases (i.e. 
it produces smaller random uncertainties but larger systematic errors). This is another consequence
of the anticorrelation between $\overline{N}_{i,{\rm min}}$ and $b$.

\begin{figure}
%\centerline{\includegraphics*[width=\linewidth]{normaldist1.ps}}
\centerline{\includegraphics*[width=\linewidth]{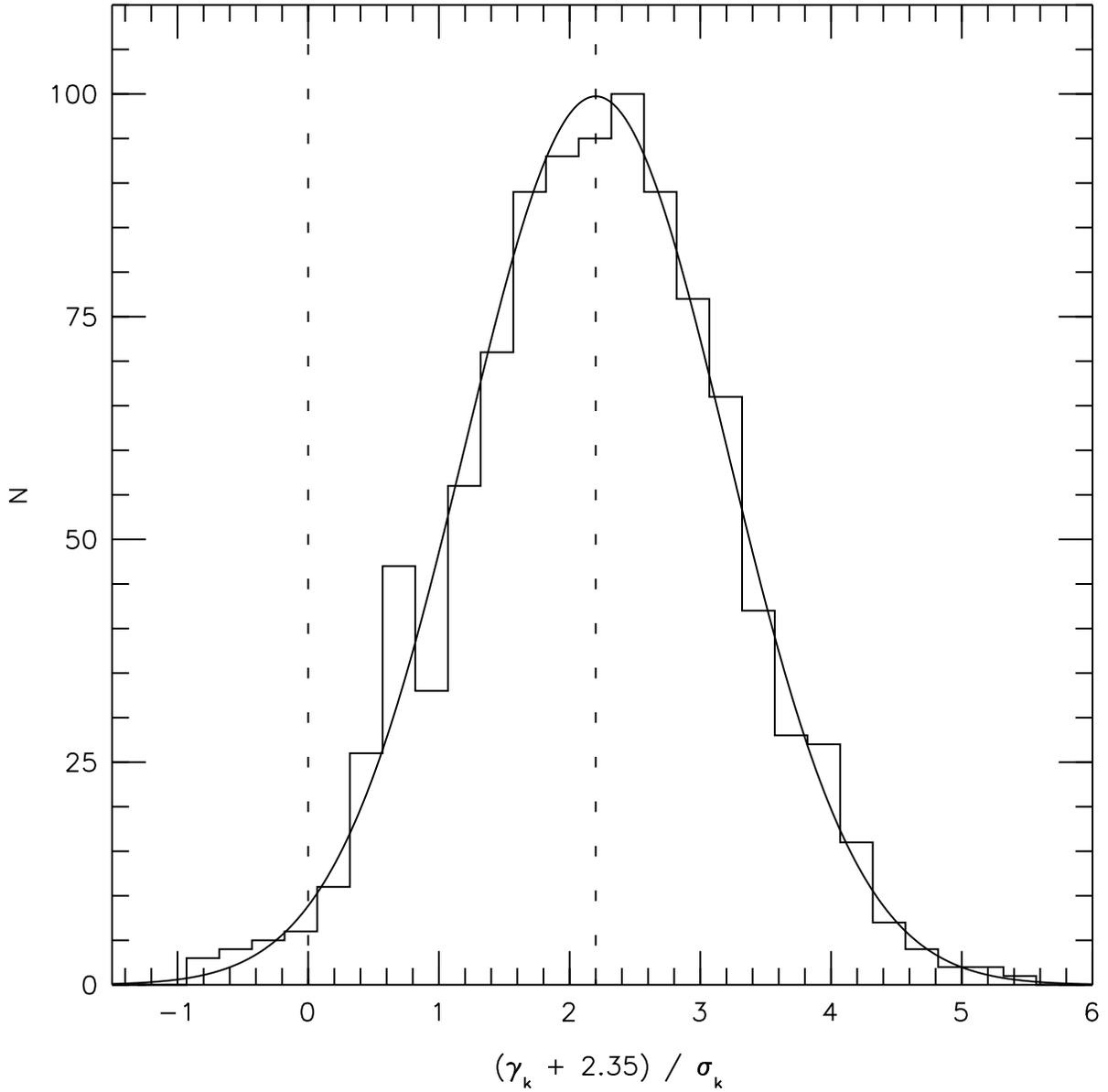}}
\caption{Histogram with the distribution of $(\gamma_k+2.35)/\sigma_k$ for the 1000 realizations of 
the first experiment with 300 stars and 50 bins. A Gaussian distribution with mean $b=2.200$ and 
dispersion of 1.0 is also plotted for comparison. The vertical lines mark the position of 0 and of $b$.}
\label{normaldist1}
\end{figure}

	We also performed a check on the values of $\beta$ and found them to be close to 1 in all cases.
Furthermore, a look at the sample case in Fig.~\ref{normaldist1} shows that the distribution of 
$(\gamma_k+2.35)/\sigma_k$ is well characterized by a Gaussian with a mean equal to $b$ and a
standard deviation of 1.0. As indicated in the previous section, this means that bins with uniform size
yield correct uncertainty estimates (i.e. the problem lies in the systematic, not in the random errors).

	We should point out that the existence of a bias towards a flattening of the IMF for
small samples when using uniform-size bins was previously detected by \citet{Krou01}, who called it
a sampling bias. As we will see in the next sections, it is possible
to get rid of such a bias almost completely.

\section{Experiment 2: variable-size bins with evenly divided number of stars per bin}

	For our second experiment we adopt the first recommendation of \citet{DAgoStep86} and use a
variable size for our bins designed in such a way as to have a similar number of stars per bin.
We do so for each of our random realizations in the following way (we use the case with 100 
stars and 5 bins as an example): 
[a] fixing $m_{\rm down}$ and $m_{\rm up}$ to be $10^{0.8}$~M$_\odot$ and $10^{2.2}$~M$_\odot$, respectively; 
[b] sorting the data so that $m_1 < m_2 < \ldots < m_{100}$; and [c] fixing the limits between 
bins $i$ and $i+1$ to be $0.5*(m_{20*i}+m_{20*i+1})$. 

\begin{deluxetable}{rrrrrrrrrrrrrrrrrrr}
\rotate
\tablecaption{Results for the variable-bin size case for 3, 5, 10, 30, and 50 bins.}
\tablewidth{0pt}
\tabletypesize{\footnotesize}
\tablehead{stars & & \multicolumn{5}{c}{$\overline{\gamma}$} & 
                   & \multicolumn{5}{c}{$\overline{\sigma}$} & 
                   & \multicolumn{5}{c}{$b$                } \\
                 & & \colhead{3} & \colhead{5} & \colhead{10} & \colhead{30} & \colhead{50} &
                   & \colhead{3} & \colhead{5} & \colhead{10} & \colhead{30} & \colhead{50} &
                   & \colhead{3} & \colhead{5} & \colhead{10} & \colhead{30} & \colhead{50}}
\startdata
   30 & &
-2.402  & -2.402  & -2.403  & -2.394  & \nodata & &    
 0.285  &  0.285  &  0.288  &  0.292  & \nodata & &    
 0.012  & -0.024  & -0.046  & -0.029  & \nodata \\     
  100 & &
-2.365  & -2.367  & -2.366  & -2.364  & -2.366  & &    
 0.152  &  0.152  &  0.154  &  0.157  &  0.156  & &    
 0.015  & -0.018  & -0.024  & -0.029  & -0.030  \\     
  300 & &
-2.353  & -2.352  & -2.353  & -2.355  & -2.354  & &    
 0.086  &  0.087  &  0.088  &  0.089  &  0.090  & &    
 0.034  &  0.031  &  0.012  & -0.012  & -0.009  \\     
 1000 & &
-2.350  & -2.348  & -2.349  & -2.349  & -2.349  & &    
 0.047  &  0.047  &  0.048  &  0.049  &  0.049  & &    
 0.043  &  0.062  &  0.045  &  0.037  &  0.043  \\     
\enddata
\label{variablebin}
\end{deluxetable}

	We show in Table~\ref{variablebin} the values of $\overline{\gamma}$, $\overline{\sigma}$,
and $b$ for the second experiment. Only nineteen cases were used because in one circumstance there 
were more bins than stars. $\overline{\gamma}$ is very close to $-$2.35 in all cases, with
a minimum of $-$2.403 and a maximum of $-$2.348. $|b|\ll 1$ in all nineteen cases (maximum value of 
0.062), with $b$ being positive in some cases and negative in others. Note that the signs of
$\overline{\gamma}+2.35$ and of $b$ can be different due to the possible existence of correlations
between the values of $\gamma_k$ and $\sigma_k$ (see also Fig.~\ref{exp3003005}). 

	These results indicate that using a variable bin size to include a similar number of stars in
each bin is a good way of minimizing binning biases. A comparison with the previous experiment shows
that this is done at no significant cost of increasing $\overline{\sigma}$. Regarding the second recommendation of 
\citet{DAgoStep86}, we only find a weak dependence of $b$ in the number of bins. The robustness of
the method is emphasized by the fact that even when 30 stars are divided into 30 bins (i.e. 1 star/bin)
no significant biases are detected. This result corroborates that the existence of large biases 
originates in the assignment of incorrect weights to each bin and not so much by the fact that 
one specific bin has a low number of stars \citep{Wheaetal95}. 

\section{Experiment 3: setting the lower and upper mass limits from the data}

	The previous two experiments have an artificial component in them: we are using as values 
for $m_{\rm down}$ and $m_{\rm up}$ the input ones, i.e., the values that are attained only in a 
sample with an infinite number of stars. When using real data, however, those values have to be
determined. An observer typically fixes the first one using incompleteness criteria (since lower-mass 
stars usually exist but are harder to detect) and the second one is usually unknown (and, likely, a
quantity one is interested in measuring, see e.g. \citealt{OeyClar05}). Therefore, we can simulate 
more realistic conditions by modifying the extremes of our second experiment by setting  
$m_{\rm down} = m_1-0.5*(m_2-m_1)$, $m_{\rm up} = m_N+0.5*(m_N-m_{N-1})$, which can be determined 
directly from the data.

\begin{deluxetable}{rrrrrrrrrrrrrrrrrrr}
\rotate
\tablecaption{Results for the variable-bin size case with data-determined $m_{\rm down}$ and 
$m_{\rm up}$ for 3, 5, 10, 30, and 50 bins.}
\tablewidth{0pt}
\tabletypesize{\footnotesize}
\tablehead{stars & & \multicolumn{5}{c}{$\overline{\gamma}$} & 
                   & \multicolumn{5}{c}{$\overline{\sigma}$} & 
                   & \multicolumn{5}{c}{$b$                } \\
                 & & \colhead{3} & \colhead{5} & \colhead{10} & \colhead{30} & \colhead{50} &
                   & \colhead{3} & \colhead{5} & \colhead{10} & \colhead{30} & \colhead{50} &
                   & \colhead{3} & \colhead{5} & \colhead{10} & \colhead{30} & \colhead{50}}
\startdata
   30 & &
 -2.365  & -2.349  & -2.329  & -2.298  & \nodata & &    
  0.308  &  0.306  &  0.305  &  0.302  & \nodata & &    
  0.110  &  0.134  &  0.180  &  0.264  & \nodata \\     
  100 & &
 -2.359  & -2.356  & -2.351  & -2.338  & -2.335  & &    
  0.155  &  0.156  &  0.157  &  0.157  &  0.158  & &    
  0.047  &  0.046  &  0.071  &  0.143  &  0.161  \\     
  300 & &
 -2.351  & -2.349  & -2.348  & -2.348  & -2.346  & &    
  0.087  &  0.088  &  0.089  &  0.090  &  0.090  & &    
  0.053  &  0.066  &  0.065  &  0.065  &  0.080  \\     
 1000 & &
 -2.349  & -2.348  & -2.348  & -2.348  & -2.347  & &    
  0.047  &  0.048  &  0.048  &  0.049  &  0.049  & &    
  0.044  &  0.079  &  0.079  &  0.073  &  0.086  \\     
\enddata
\label{variablebinedge}
\end{deluxetable}

\begin{figure}
%\centerline{\includegraphics*[width=\linewidth]{biascorr3.ps}}
\centerline{\includegraphics*[width=\linewidth]{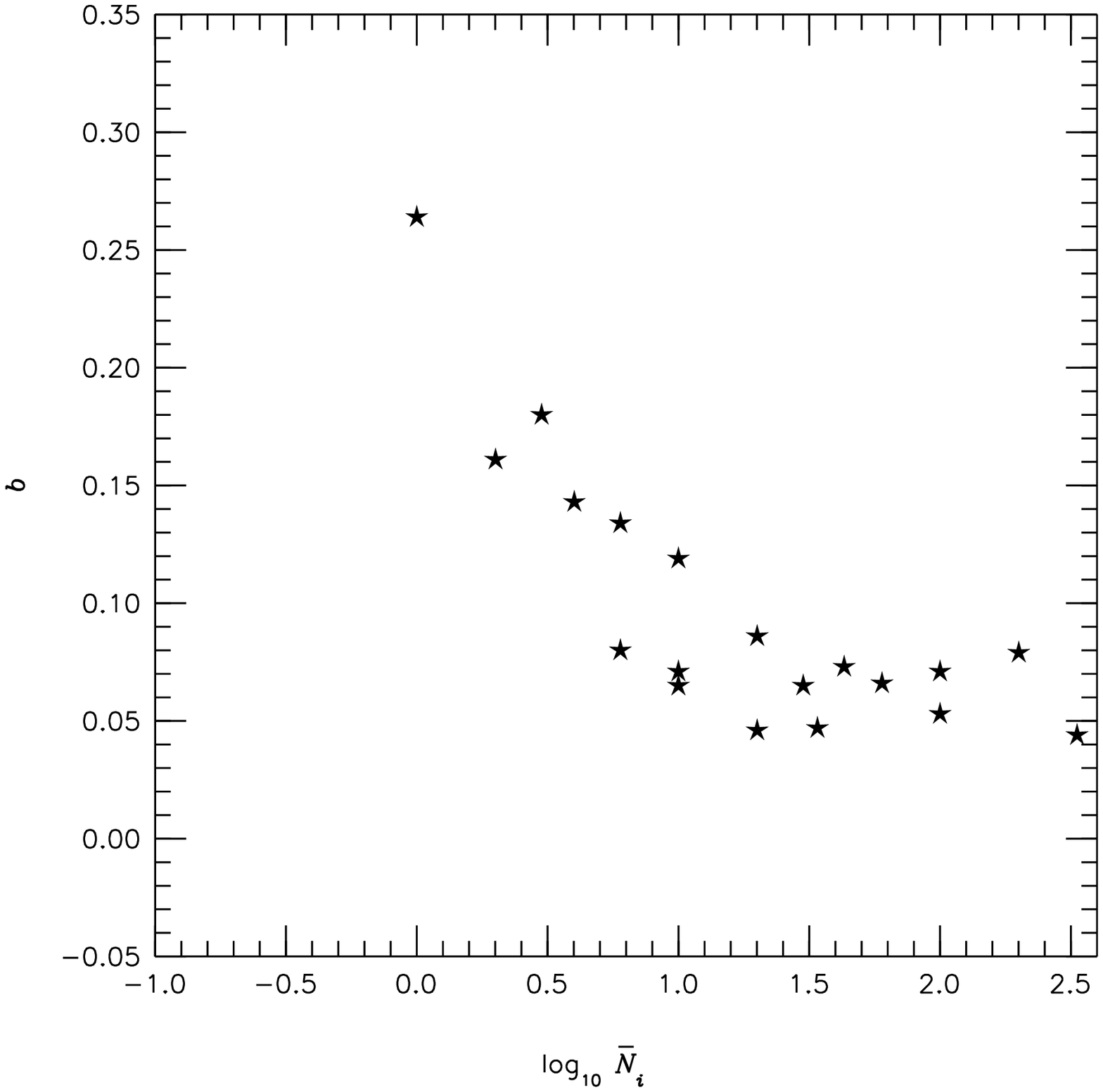}}
\caption{Bias as a function of $\overline{N}_{i}$ for the third experiment. Note that the vertical
scale for the plot is 1/10 that of Fig.~\ref{biascorr1}.}
\label{biascorr3}
\end{figure}

	We show in Table~\ref{variablebinedge} the values of $\overline{\gamma}$, $\overline{\sigma}$,
and $b$ for the third experiment. Results are similar to the ones for the second experiment, with 
values of $\overline{\gamma}$ between $-$2.365 and $-$2.298. This is especially so for the cases with
300 and 1000 stars, as expected. Comparing results one by one, we find that
$\overline{\gamma}$ is always larger here than for the previous experiment. $b$ is now always positive
and the values of $|b|$ are somewhat higher than in the previous experiment. Still, all cases with more 
than 3 stars per bin have $|b|\le 0.134$ and even the worst case has only $|b| = 0.264$. In 
Fig.~\ref{biascorr3} we see that $b$ is still anticorrelated with $\overline{N}_i$, the average number of 
stars per bin.

	These results indicate that fixing the lower and upper mass limits from the data and using a 
variable bin size to include a similar number of stars in each bin is a practical way of 
minimizing binning biases. There is a price in the former of larger biases that has to be paid for the 
lack of knowledge of the ends of the distribution, but that price is small, as one can observe by 
comparing Figs.~\ref{biascorr1}~and~\ref{biascorr3}. We should point out that the second
recommendation of \citet{DAgoStep86} does not work out for this experiment: if one is interested in 
minimizing biases, then it is preferable to go with a low number of bins (but, as previously indicated,
biases are never large anyway).

\begin{figure}
%\centerline{\includegraphics*[width=\linewidth]{normaldist3.ps}}
\centerline{\includegraphics*[width=\linewidth]{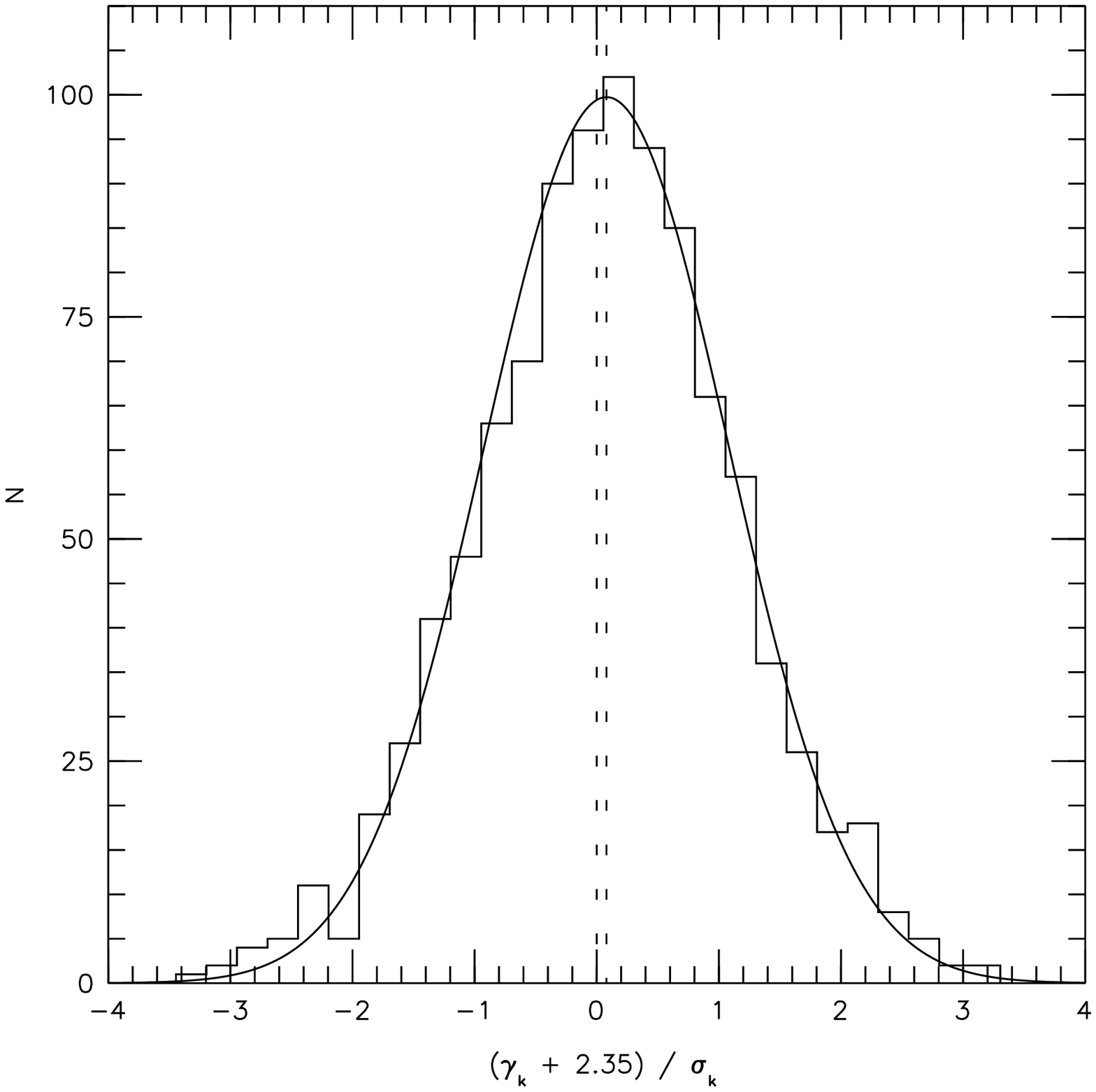}}
\caption{Histogram with the distribution of $(\gamma_k+2.35)/\sigma_k$ for the 1000 realizations of 
the third experiment with 300 stars and 50 bins. A Gaussian distribution with mean $b=0.090$ and 
dispersion of 1.0 is also plotted for comparison. The vertical lines mark the position of 0 and of $b$.}
\label{normaldist3}
\end{figure}

	The values of $\beta$ are found to be close to 1 in all cases and the histogram for the sample
case in Fig.~\ref{normaldist3} shows the same behavior as the one in Fig.~\ref{normaldist1}: the distribution of 
$(\gamma_k+2.35)/\sigma_k$ is well characterized by a Gaussian with a mean equal to $b$ (and, in this
case, close to zero) and a dispersion of 1.0. Therefore, {\it our technique yields not only a nearly bias-free
value for the slope of the IMF but also a correct estimate of its uncertainty.}

\begin{figure}
%\centerline{\includegraphics*[width=0.35\linewidth]{exp3_0030_05_a.ps}}
%\centerline{\includegraphics*[width=0.35\linewidth]{exp3_0030_05_b.ps}}
%\centerline{\includegraphics*[width=0.35\linewidth]{exp3_0030_05_c.ps}}
\centerline{\includegraphics*[width=0.35\linewidth]{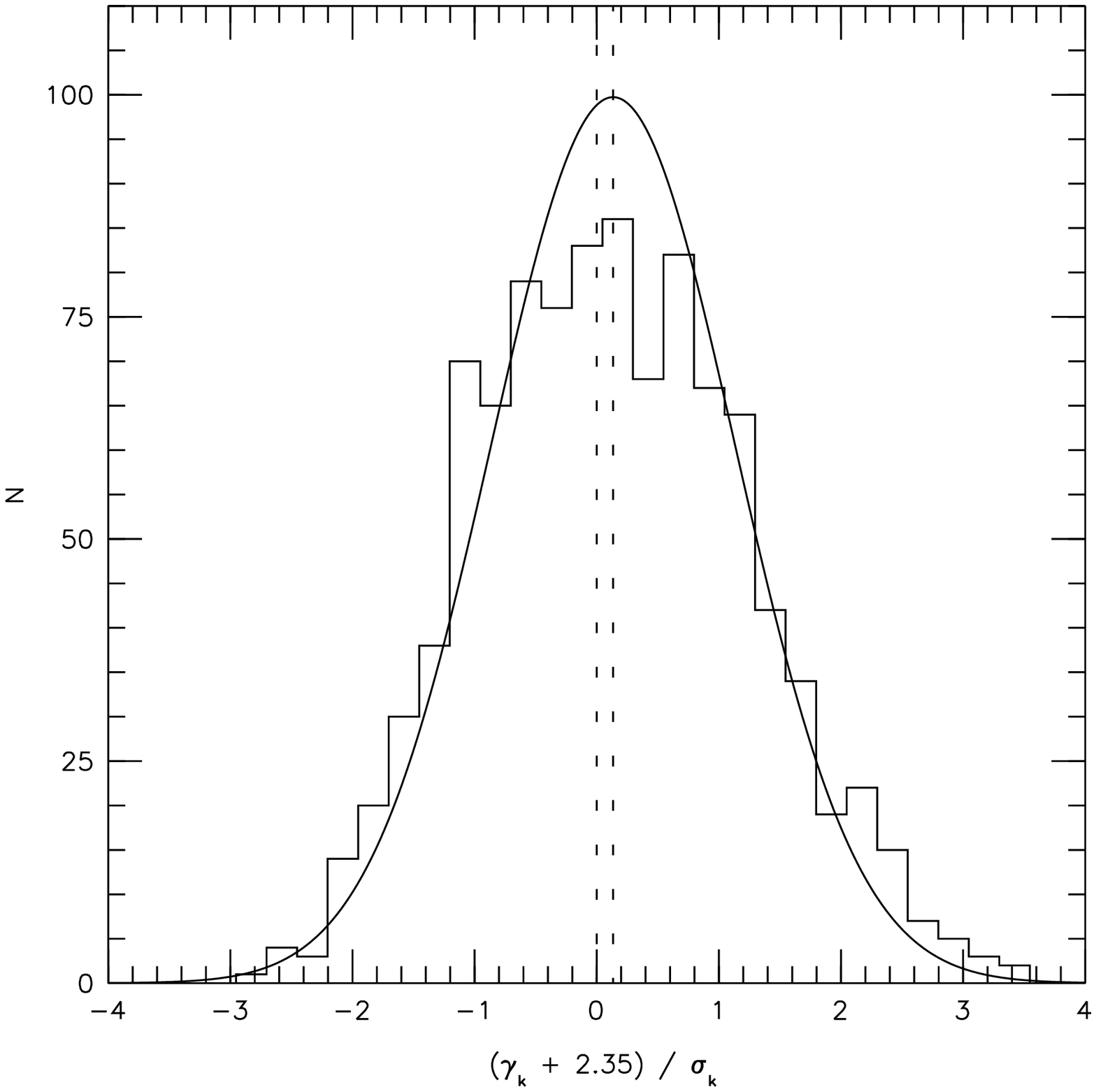}}
\centerline{\includegraphics*[width=0.35\linewidth]{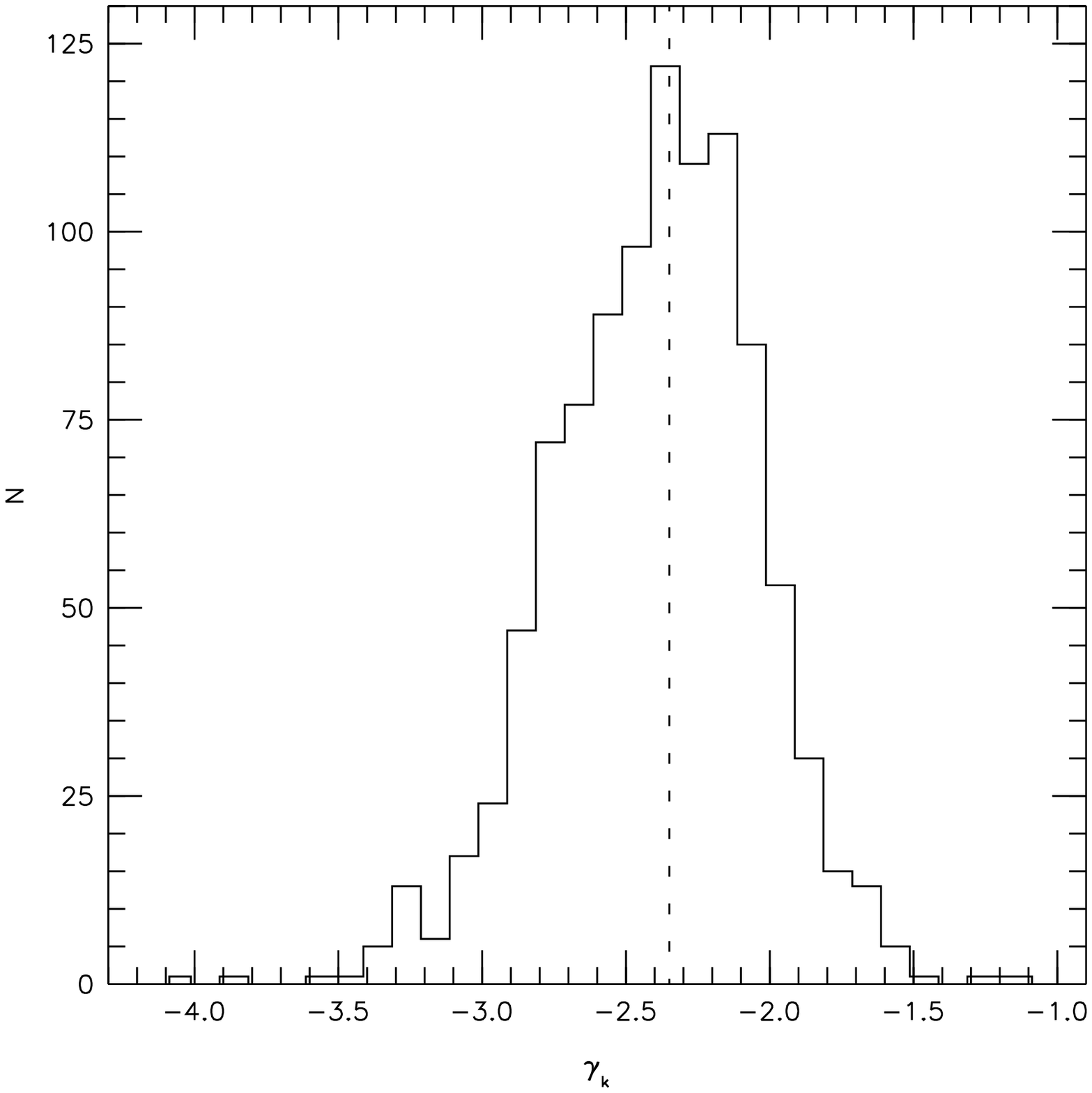}}
\centerline{\includegraphics*[width=0.35\linewidth]{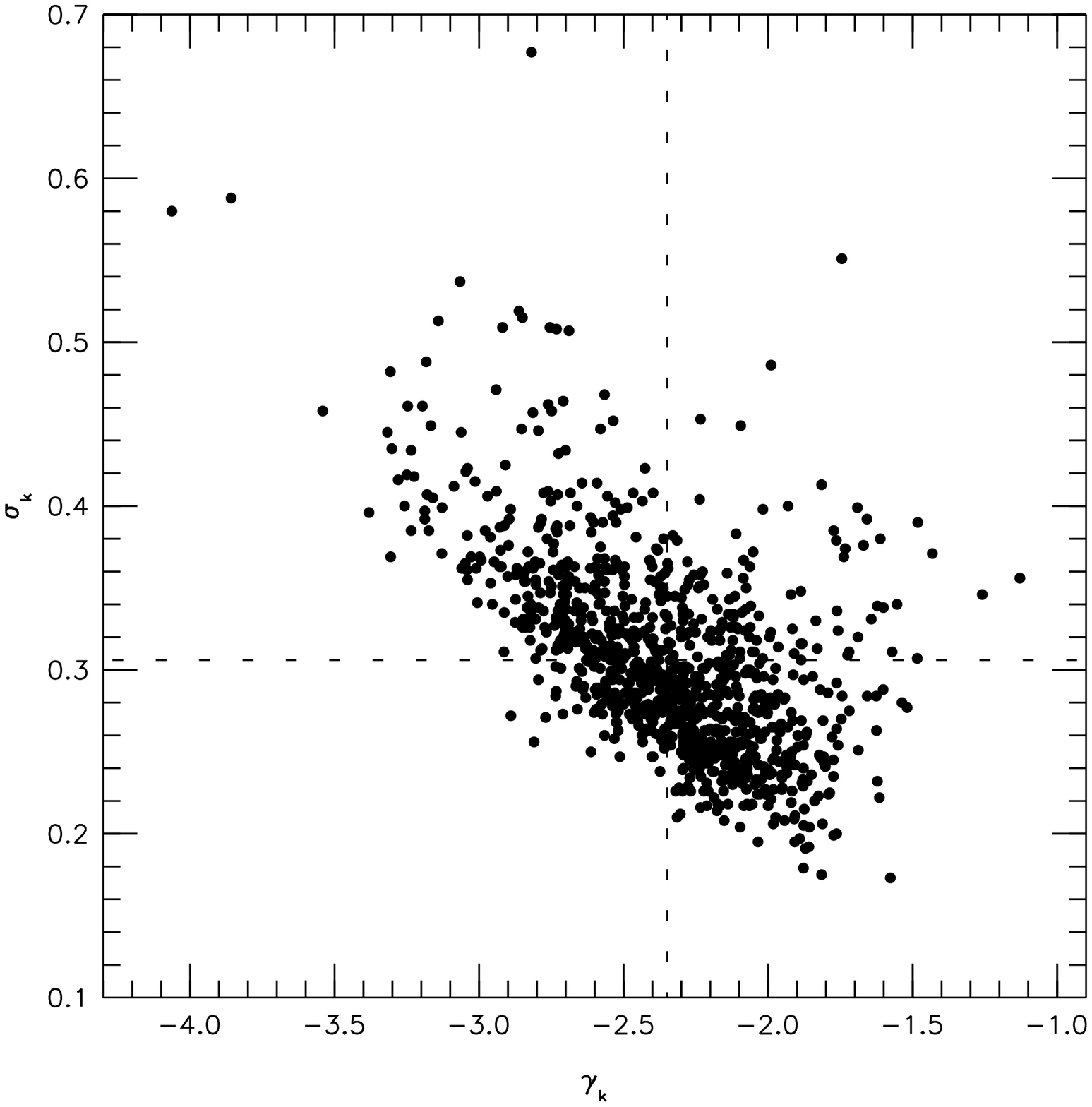}}
\caption{Detailed results for the 30 stars + 5 bins case for the third experiment. (top) Histogram with the 
distribution of $(\gamma_k+2.35)/\sigma_k$. A Gaussian distribution with mean $b=0.134$ and 
dispersion of 1.0 is also plotted for comparison. Vertical lines mark the position of 0 and $b$.
(center) Histogram with the distribution of $\gamma_k$. A vertical line marks the position of
$\overline{\gamma}$. (bottom) Results for $\gamma_k$ and $\sigma_k$ for the 1000 realizations. Lines
mark the values of $\overline{\gamma}$ and $\overline{\sigma}$.}
\label{exp3003005}
\end{figure}

	Another advantage of the method proposed 
here can be extracted from Table~\ref{variablebinedge}: a nearly bias-free measurement of the power-law 
exponent with an uncertainty of less than 0.2 can be obtained with only 100 stars. If the number is 
lowered to 30 stars, then the uncertainty in the power-law exponent is close to 0.3. 
Detailed results for the 30 stars + 5 bins case are shown in Fig.~\ref{exp3003005}.
Note how the first histogram shows a symmetric distribution 
while the second one is distinctly asymmetric. The difference is explained by the
correlation ($r=-0.58667$) between $\gamma_k$ and $\sigma_k$ shown in the bottom plot: chi-square fitting
yields larger values of the uncertainty in the slope for lower values of the slope itself. Note that
this correlation is not a problem in itself because the histogram that determines through its mean and
dispersion whether the technique yields a correct estimate of the IMF slope is the first one, not the second. 
Therefore, we conclude that {\it it is possible to conduct precise studies of the mass segregation within 
a large cluster or to measure the IMF in a small one.} 

\begin{figure}
%\centerline{\includegraphics*[width=0.375\linewidth]{comparehistoa.ps}}
%\centerline{\includegraphics*[width=0.375\linewidth]{comparehistob.ps}}
%\centerline{\includegraphics*[width=0.375\linewidth]{comparehistoc.ps}}
\centerline{\includegraphics*[width=0.375\linewidth]{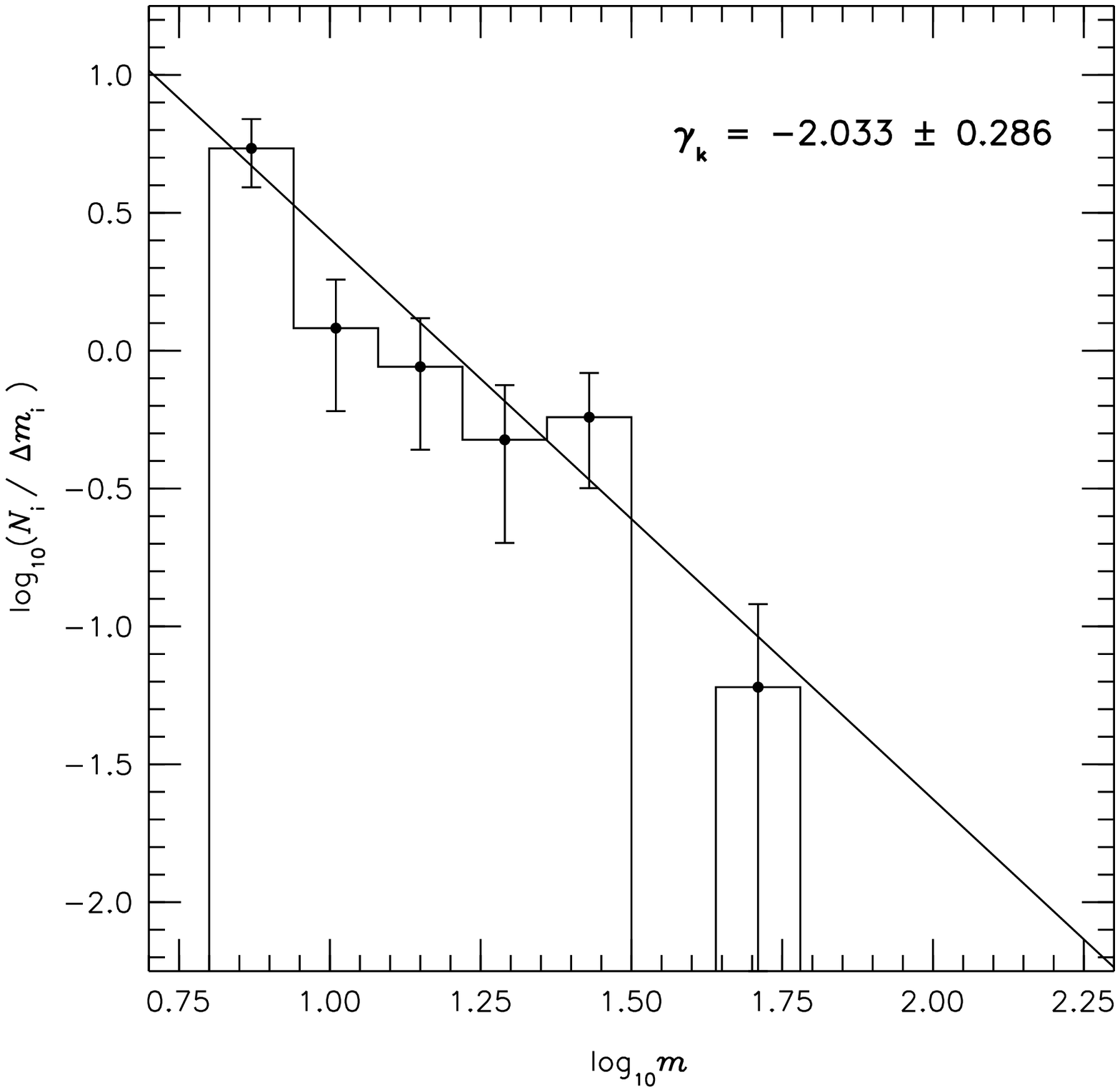}}
\centerline{\includegraphics*[width=0.375\linewidth]{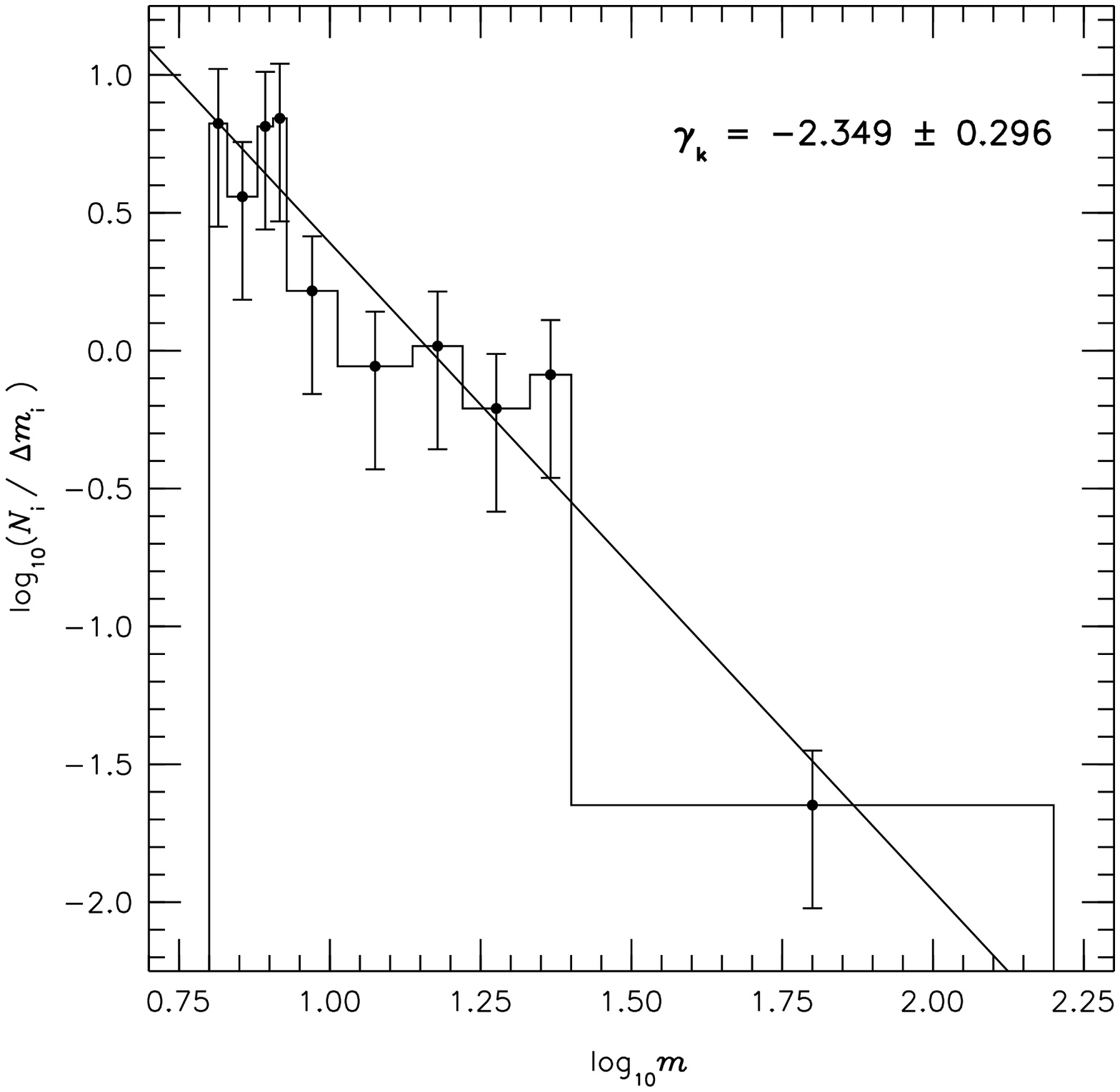}}
\centerline{\includegraphics*[width=0.375\linewidth]{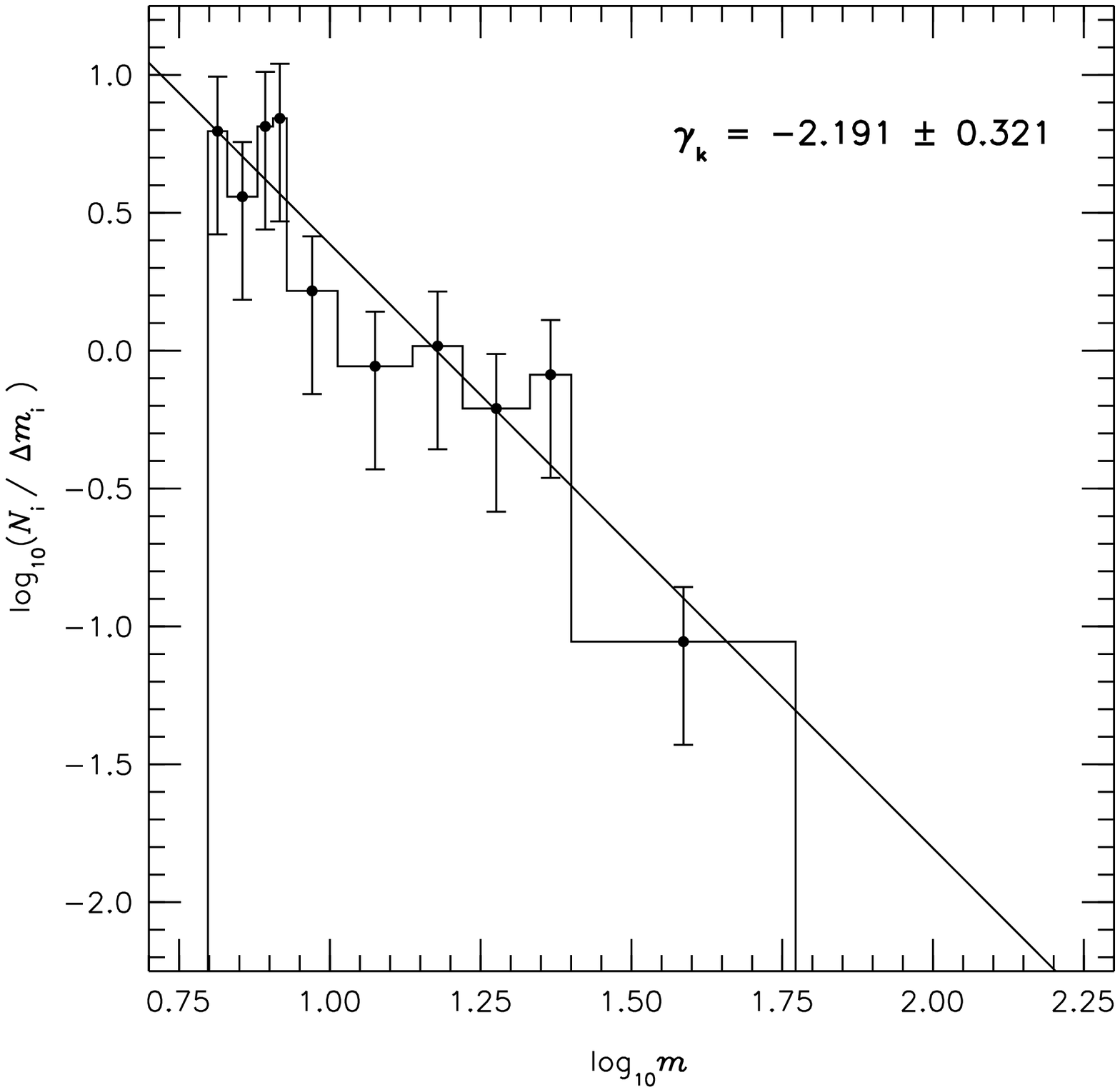}}
\caption{Comparison between the data and the fitted functions for one of the realizations with
30 stars and 10 bins for the three experiments in this article. The top, center, and bottom panels show 
the first, second, and third experiments, respectively. Note the differences in the size of the error 
bars for the histograms between the first and the last two experiments.}
\label{comparehisto}
\end{figure}

	Different binning schemes can easily yield different values for the same sample. This can be seen in
Fig.~\ref{comparehisto}, where we show a comparison between the three experiments for a single realization. 
With that in mind, one wonders whether part of the variations in the IMF detected by a number of authors
(see \citealt{Elme04} for a recent review) are not real but simply numerical effects introduced by the 
different schemes used. In order to test that, one would have to reanalyze the data in a uniform manner 
using an unbiased scheme, such as the one presented in this article.

\section{Summary and future work}

	We conclude that the binning mechanism proposed in this paper for the fitting of power laws
with Salpeter slopes yields results that (a) are nearly bias-free and (b) produce correct uncertainty 
estimates, as tested by our numerical simulations. On the other hand, the standard uniform-size binning
introduces biases that are dependent on the number of stars per bin. The power of the technique described
here extends to small samples, since we have shown that it is possible to obtain accurate values
with reasonable precisions for the IMF slope even when as few as 30 stars are available for analysis.

	We are finishing an analysis of HST/WFPC2 stellar photometry of the nearby dwarf starburst galaxy 
NGC 4214. In that article we will apply the technique described here in order to study the IMF for the 
massive stars in that galaxy. We will also investigate other possible
sources of biases in the calculation of the IMF.

	We would also like to point out that, given the purely numerical nature of our analysis, our
results could be extended to other similar problems. For example, the mass function for young stellar
clusters can be rather well approximated by a power law with a slope of $-2.0$ (see e.g. 
\citealt{FallZhan01}), which is quite close to $-2.35$, so the same type of biases should be present 
there as well. Note, however, that for a problem where we expect the power law to have a radically 
different exponent (e.g. $-10.0$),
the results in this paper may not apply because of the larger disparity between the values of the function at
the two extremes. In general, we recommend that biases be evaluated for any function fitted to binned data
through chi-square minimization by means of specific numerical experiments similar to the ones in this 
article.

\begin{acknowledgements}

We would like to thank the referee, Miguel Cervi\~no, for very useful suggestions and for comments 
that helped improved this article considerably, especially concerning the issue of weights.
Support for this work was provided by NASA through grants GO-09419.01-A and AR-09553.02-A from the 
Space Telescope Science Institute, Inc., under NASA contract NAS5-26555.

\end{acknowledgements}

\bibliographystyle{apj}
\bibliography{general}

\end{document}